\documentclass[acmsmall]{acmart}

\usepackage{tabularx}
\usepackage{ulem} 
\usepackage{array,multirow}
\usepackage{ltablex}    
\keepXColumns           
\usepackage{multirow}   
\usepackage{caption}    

\AtBeginDocument{%
  \providecommand\BibTeX{{%
    \normalfont B\kern-0.5em{\scshape i\kern-0.25em b}\kern-0.8em\TeX}}}

\setcopyright{acmcopyright}
\copyrightyear{2025}
\acmYear{2025}
\acmDOI{}

\acmPrice{15.00}
\acmISBN{2573-0142/2025/4-ARTCSCW132}

\setcopyright{cc}
\setcctype{by}
\acmJournal{PACMHCI}
\acmYear{2025} \acmVolume{9} \acmNumber{7} \acmArticle{CSCW305}  

\acmMonth{11} \acmPrice{}

\acmDOI{10.1145/3757486}

\newcommand{\blue}[1]{\textcolor{black}{#1}}
\newcommand{\red}[1]{\textcolor{black}{#1}}
\begin{document}

\title{Interaction Configurations and Prompt Guidance in Conversational AI for Question Answering in Human-AI Teams}
\renewcommand{\shorttitle}{Interaction Configurations and Prompt Guidance in Conversational AI}
\author{Jaeyoon Song}
\affiliation{%
 \institution{Massachusetts Institute of Technology}
 \city{Cambridge}
 \state{MA}
 \country{United States}
}
\author{Zahra Ashktorab}
\affiliation{%
 \institution{IBM Research}
 \city{New York}
 \state{NY}
 \country{United States}
}
\author{Qian Pan}
\affiliation{%
 \institution{IBM Research}
 \city{Cambridge}
 \state{MA}
 \country{United States}
}
\author{Casey Dugan}
\affiliation{%
 \institution{IBM Research}
 \city{Cambridge}
 \state{MA}
 \country{United States}
}
\author{Werner Geyer}
\affiliation{%
 \institution{IBM Research}
 \city{Cambridge}
 \state{MA}
 \country{United States}
}
\author{Thomas W. Malone}
\affiliation{%
 \institution{Massachusetts Institute of Technology}
 \city{Cambridge}
 \state{MA}
 \country{United States}
}

\renewcommand{\shortauthors}{Jaeyoon Song et al.}

\received{July 2024}
\received[revised]{December 2024}
\received[accepted]{March 2025}

\begin{abstract}
Understanding the dynamics of human-AI interaction in question answering is crucial for enhancing collaborative efficiency. Extending from our initial formative study, which revealed challenges in human utilization of conversational AI support, we designed two configurations for prompt guidance: a Nudging approach, where the AI suggests potential responses for human agents, and a Highlight strategy, emphasizing crucial parts of reference documents to aid human responses. Through two controlled experiments, the first involving 31 participants and the second involving 106 participants, we compared these configurations against traditional human-only approaches, both with and without AI assistance. Our findings suggest that effective human-AI collaboration can enhance response quality, though merely combining human and AI efforts does not ensure improved outcomes. In particular, the Nudging configuration was shown to help improve the quality of the output when compared to AI alone. This paper delves into the development of these prompt guidance paradigms, offering insights for refining human-AI collaborations in conversational question-answering contexts and contributing to a broader understanding of human perceptions and expectations in AI partnerships.
\end{abstract}




\maketitle

\section{Introduction}

As more artificial intelligence (AI) models achieve high performance, their adoption across various domains \cite{wang2019human, wang2020human, park2019identifying} facilitates user task completion and decision-making. Determining the most effective mode of interaction for optimizing outcomes within the resulting human-AI partnership is not always straightforward \cite{xu2021human, xu2023transitioning, kambhampati2020challenges}. With the widespread utilization of large language models (LLMs) capable of producing highly sophisticated text, the question of the most suitable mode or style of interaction in human-AI collaboration becomes even more complex, as the interaction is not just decision-making but also reviewing aspects of the AI-generated text, editing, improving it, and appending additional information to it. Further, generating textual responses in a question-answering context brings with it requirements in tone (i.e. likely polite) and correctness, which differentiates it from other kinds of text-generation tasks such as creative writing. In the middle of such complexity, little is known about how different interfaces and configurations of interaction can potentially impact how people create text using LLMs in this context. In this work, we aim to find out how we can aid humans in better interacting with LLMs to generate texts, specifically in the context of question answering. In particular, we simulate and test materials that would be available in a customer support environment, the representative question-answering application in the real world.

\begin{figure}[h]
    \caption{Experiment Design and Study Conditions for Study 1}
    \label{fig:study1_design}
    \centering\includegraphics[width=0.9\textwidth]{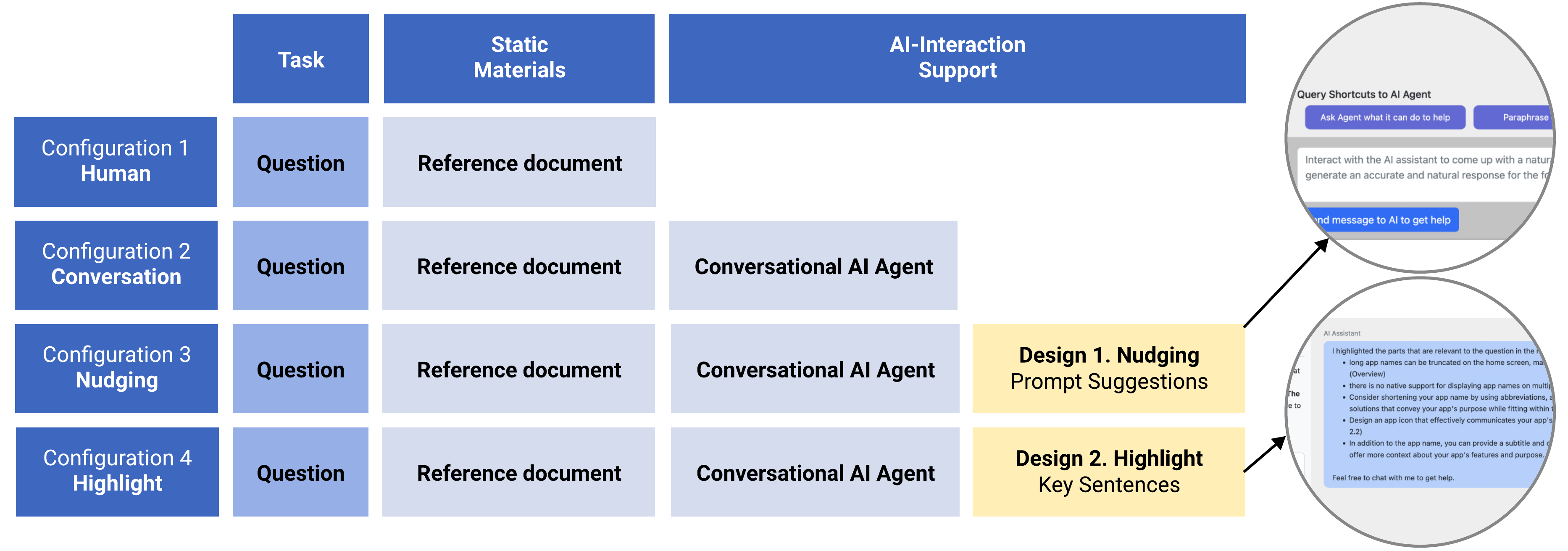}
\end{figure}

Human-AI collaboration in question answering combines the efficiency and speed of automation with human interaction, potentially resulting in improved response quality and enhanced satisfaction of the inquirer \cite{jacobsen2020perceived, liu2022artificial, xu2020ai, adam2021ai}. Although prior research has suggested the significance and potential of employing humans and AIs together for question answering~\cite{amershi2019guidelines, campero2022test}, little has been studied about what leads to successful human-AI interaction, particularly between LLMs and humans. In this paper, we test two configurations that provide guidance for human agents using conversational AI for question answering. We compare these configurations against traditional human-only approaches, both with and without AI assistance. We also examine what factors correlate with the quality of the responses and lead to successful human-AI collaboration and what is \textit{perceived} as successful high-quality responses. Based on our findings, we suggest several considerations that researchers and practitioners should be mindful of when designing human-AI interaction in question answering. These considerations can further be extended to the customer support context that inherently involves question-answering. 

To inform our experiment, we first performed a formative study to explore four types of human-AI collaboration. All four configurations shared three basic elements: a \textit{question} to respond to, a \textit{reference document} that contains accurate information needed to answer the question, and access to a \textit{conversational AI agent} that is fed with the question and reference document, powered by \blue{GPT-4}. Then we asked participants to construct an accurate response that can satisfy the given question, under each configuration. Based on the findings from the formative study, we designed two configurations that can possibly cope with the identified challenges: the \textit{Nudging} configuration that suggests messages to send to the AI agent and the \textit{Highlight} configuration that presents key sentences of the reference document.

To evaluate the effectiveness of these two designs and investigate factors that influence human-AI interaction, we conducted two controlled experiments. In the first experiment (Study 1), we asked participants to construct an accurate response to the question, similar to the formative study. We compared four configurations including the Nudging and Highlight conditions. The other two configurations served as baselines for comparison purposes: the Human condition without AI and the Conversation condition that provides only the AI agent without additional support. All four conditions shared a question and a reference document. Figure~\ref{fig:study1_design} summarizes the design of Study 1. Then in a follow-up experiment (Study 2), a separate group of raters was recruited to evaluate the responses collected in Study 1, along with responses constructed by AI alone. The raters were first trained with several examples and then performed pairwise comparisons on the responses. Results show that human-AI collaboration is potentially useful, but only when their interaction is successful. The mere combination of humans and AI does not automatically guarantee performance improvement. When considering the overall quality, there was no significant difference in the quality of responses generated between any of the conditions. However, when considering the best seven responses in each condition, responses constructed in the \textit{Nudging} and \textit{Conversation} conditions were significantly better than those generated by AI alone, but not human alone. 

Several insights on how to design human-AI collaboration in question answering emerged from our analysis. First, it is desirable and necessary to encourage more active human engagement in the collaboration process. Second, the message suggestions were shown to improve the interaction. For instance, the \textit{Nudging} condition had more successful responses from the AI than the \textit{Highlight} or \textit{Conversation} conditions. Third, inducing humans to ask AI what to ask AI is helpful. This kind of meta-prompting was shown to be positively correlated with the number of successful AI responses. Finally, the design should ensure that the humans and AI use a similar vocabulary. Although paraphrasing a question was positively correlated with the number of successful AI responses, paraphrasing sometimes caused failures on the AI side. 

The contributions of this work are as follows:

\begin{itemize}
\item The design of prompt guidance configurations informed by our formative study which empowers users to leverage AI assistance to accomplish a task. 
\item Findings from the two main studies that showed responses generated in the AI-only condition were ranked higher than those in the human-only and human-AI combination conditions. However, when considering only the top-rated responses from each condition, the quality of responses in the human-AI condition surpassed the others.
\item A design recommendation that message suggestions can enhance interaction, with the Nudging condition resulting in more successful AI responses compared to the Highlight or Conversation conditions.
\end{itemize}


\section{Related Work}

\subsection{Human-AI Collaboration} 
When humans collaborate with artificial intelligence (AI) towards a shared goal, the expectation is that their combined abilities will lead to enhanced performance and achievement beyond what either party could accomplish individually. Evaluation of such human-AI teams often revolves around performance metrics or subjective satisfaction of the human participants \cite{capel2023human}. Human-AI collaboration takes diverse forms and spans numerous domains. In the context of healthcare, for example, collaboration between humans and AI can be observed in patient-centered healthcare, online health communities, and mental health treatment~\cite{lai2021human, park2019identifying, thieme2023designing}. Research in the medical domain by Reverberi et al. \cite{reverberi2022experimental} explored the interaction between medical doctors (MDs) and AI. The study found that endoscopists were influenced by AI suggestions but followed the advice more consistently when it was correct. Similarly, Baniecki et al. \cite{baniecki2023hospital} investigated the prediction of a patient's length of stay in a hospital using only an X-ray image and introduced time-dependent model explanations into the human-AI decision-making process.

Prior research has explored multiple factors that can impact the quality of the collaboration between the human and an AI agent.  Inkpen et al.~\cite{inkpen2022advancing} examined users’ interactions with three simulated algorithmic models, all with equivalent accuracy rates but each tuned differently in terms of true positive and true negative rates. They found that several underlying factors, including user base expertise and complementary human-AI tuning, significantly impact the overall team performance. Holstein et al.~\cite{holstein2023toward} conducted an online experiment to understand whether and how explicitly communicating potentially relevant unobservables influences how people integrate model outputs and unobservables when making predictions. Xu et al.~\cite{xu2023comparing} investigated the tradeoff between precision and recall in the AI’s recommendations in a real-world video anonymization task. They analyzed the performance of professional annotators working with different types of AI assistance. Munyaka et al.~\cite{munyaka2023decision} investigated how the decision-making style of a team member in a human-AI team impacts the outcome of the collaboration and perceived team efficacy. They found significant differences across different decision-making styles and disclosed AI identity disclosure in an AI-driven collaborative game. Cabrera et al.~\cite{cabrera2023improving} proposed showing users behavior descriptions, and details of how AI systems perform on subgroups of instances, to help people appropriately rely on AI aids. They found that behavior descriptions can increase human-AI accuracy by helping people identify AI failures and increasing people's reliance on the AI when it is more accurate.

Collaboration between humans and AI is crucial in high-stakes domains such as policing \cite{busuioc2021accountable, dakalbab2022artificial}, recidivism \cite{hillman2019use}, and child welfare \cite{saxena2020human}. AI can process vast amounts of data to identify patterns and trends, while humans provide essential context and ethical judgment. In policing, AI aids in crime prediction \cite{busuioc2021accountable, dakalbab2022artificial}, but human interpretation ensures respect for individual rights. In recidivism \cite{hillman2019use}, AI identifies risk factors, while human professionals apply these insights with an understanding of individual circumstances. In child welfare, AI helps detect patterns of neglect, but human social workers interpret these findings considering family dynamics. The collaboration between humans and AI not only leverages the strengths of both but also serves as a check and balance, enhancing decision-making and mitigating the risk of unjust outcomes due to AI errors.

In creative fields such as music \cite{mccormack2019silent} and storytelling \cite{shakeri2021saga,mirowski2023co},  the collaboration between humans and artificial intelligence (AI) is fostering innovation. AI algorithms generate novel elements in music \cite{mccormack2019silent}, providing inspiration for composers and musicians, while human artists infuse these creations with emotion and cultural context. In storytelling, AI serves as a tool for writers, generating plot ideas and character descriptions, but it's the human touch that shapes these into compelling narratives \cite{shakeri2021saga}. In these domains, the collaboration between humans and AI sparks a creative dialogue, pushing the boundaries of what's possible.


As conversational AI agents continue to gain popularity and their capabilities expand, extensive research efforts have been dedicated to enhancing their design. Prior to the widespread adoption of large language models (LLMs), a multitude of studies in HCI communities focused on comprehending user interactions with supervised intent-based classifiers. These studies explored various aspects such as personalization techniques \cite{joshi2017personalization}, repairing breakdowns in communication \cite{ashktorab2019resilient}, and the utilization of visual cues like embodiment \cite{pampouchidou2017automatic}. There also has been research on query suggestions that help users prompt LLM~\cite{kelly2009comparison, yan2018smarter} as well as similar recommendation features currently available in popular LLM user interfaces such as ChatGPT and Bing Chat. While many insights from the aforementioned HCI research can still be applicable to LLM-driven conversational agents, it is crucial to conduct additional studies to understand how users interact with these advanced tools. As LLMs introduce new complexities and possibilities, researchers must explore how users perceive and engage with these AI agents to ensure the best possible performance and outcomes. By gaining a deeper understanding of user-agent interactions in the context of LLM-driven conversational agents, researchers can develop guidelines and provide appropriate guidance to users, ultimately improving the usability and effectiveness of these AI systems \cite{amershi2019guidelines}. 

\subsection{AI-Assisted Question Answering}

Question answering represents a fundamental task in natural language processing (NLP)~\cite{rajpurkar2018know, rajpurkar2016squad, choi2018quac}. Extractive question answering, or reading comprehension, is a basic type of question answering where a model responds to a question based on a provided context~\cite{kwiatkowski2019natural, kim2021linguist}. More advanced question-answering tasks involve generating open-ended responses from a given context or even from a broad, open domain. While significant strides have been made in automated question-answering research, current systems often fall short by lacking humane qualities such as creativity and natural tone, and they do not consistently achieve perfect accuracy~\cite{molla2007question, allam2012question, diekema2004evaluation}. Therefore, leveraging human-AI collaboration can prove invaluable in enhancing question-answering capabilities.

Numerous researchers have scrutinized data from question-answering platforms such as StackOverflow and Stack Exchange to uncover their dynamics. Chua \& Banerjee~\cite{chua2015answers} validated the Quest-for-Answer framework through an empirical case study of Stack Overflow, focusing on answerability. Kabir et al~\cite{kabir2024stack} performed a comprehensive linguistic analysis of ChatGPT responses, examining both linguistic features and human aspects. Tian et al.~\cite{tian2013towards} highlighted the critical role of contextual information in achieving successful answers within community-based questioning services. Recent research in AI-assisted question answering has delved into various aspects, aiming to enhance the quality and speed of responding to the questions. One area of investigation focused on the potential of AI chatbots in facilitating positive change and support for inquirers in the digital realm \cite{toader2019effect}. This research explored the psychological responses of inquirers when interacting with virtual assistants, emphasizing the importance of understanding the perspectives of the inquirers.


Recent research in AI-assisted customer support has explored themes such as understanding customer perspectives, intent discovery, adoption factors, and customer experience enhancement \cite{nicolescu2022human}. By investigating these areas, researchers aim to improve the overall effectiveness and satisfaction of customer service interactions. With the proliferation of conversational agents driven by large language models, we build on this prior work to understand how individuals interact with conversational agents to answer a potential customer query and ultimately which configurations of human-AI interactions lead to responses most preferred by users.

\section{Basic Components of the Study}
\label{sec:basic}

In our formative and main studies, we provided three basic elements in the task: a \textit{question} to respond to, a \textit{reference document} that contains accurate information needed to answer the question, and a \textit{conversational AI agent} that is fed with the question and reference document. These elements comprise the question-answering environment in our formative study and experiments. In particular, we wanted to simulate materials that would be available in a customer support environment, the representative question-answering application in the real world. The task for participants was to construct an accurate response that answers the given question, based on the reference document. In this section, we describe how we came up with the questions and reference documents.

\begin{table}[h]
\caption{Questions presented to participants in Study 1}
\small
\label{tab:questions}
\begin{tabularx}{\textwidth}{|l|p{4cm}|X|}
\hline
Index & Title & Body \\
\hline
1     & iphone voice memos 'ping' sound                        & I'm writing a voice recording application, and I'd like to play the ping-ping sound that the voice memos program plays when it finishes recording.  Is there any way to accomplish this?  If not, does anyone know where I can find the sound? Thanks! \\
\hline
2     & What is App store screenshot size for 6.5" display?    & Apparently, Apple documentation can't keep up with App Store changes. Until today, the biggest (optional) display size was a 5.8-Inch Super Retina Display with a resolution of 1125 x 2436 pixels. Currently, it is a 6.5-inch display, but the resolution is nowhere to be found (docs at https://help.apple.com/app-store-connect/\#/devd274dd925 only mention 5.8-inch display). App store connect doesn't mention the resolution and error message "The dimensions of one or more screenshots are wrong. Learn More." redirects to docs linked above. \\
\hline
3     & In app purchase does not work when live                & I have an app in the app store that I have added an in app purchase for, I fully tested it in the sandbox environment and all worked fine. I have had the app update and the in app purchase approved by apple and according to iTunes connect all has gone live. The app updates just fine but then in app purchase simply doesn't work!! No products are returned! Anyone else have this problem? Surely if it worked in the sandbox and apple has approved it, there should be no issues!? \\
\hline
4     & How I know which contacts have iMessage?               & I have iPhone 6. Is there any app on apple app store that can show me which of my contacts have IOS iMessage? There is some information on google but for checking one by one contact. I want to check all contacts in one shot  \\
\hline
5     & Apple Health: export health data automatically         & I am trying to export the health data from my iPhone/Apple Watch automatically. I know it is possible to export the data manually by opening the Health app on your iPhone clicking the user icon "Export Health Data". However, I would like to do this automatically and periodically to analyze the data (heart rate, steps, etc.) externally in Python. Is this possible at all? \\
\hline
6     & Long iPhone App name to be displayed on multiple lines & My App name is 17 characters long. When installed on a device it looks like App...Name. Is there any way to display app names on multiple lines? Please help.\\
\hline
7     & Installing Android Studio app on Iphone for testing    & Is there a way I can install Android on an iPhone for testing purposes? If so, could I move my app from Android Studio to my iPhone then? I was looking at Andrios which is on Cydia (for Jailbroken iPhones) \\ 
\hline
\end{tabularx}
\end{table}

\subsection{Questions}

We collected question-and-answer sets from Stack Overflow as the data in our studies. We chose Stack Overflow because it is one of the most active question-answering platforms online. To ensure relevance to a typical question-answering setting, we selected questions about iPhone usage and deployment. We chose iPhone usage and deployment as the topic because they are one of the least technical topics on the Stack Overflow platform, which makes it more generalizable to other contexts. We excluded questions with code-heavy tags (e.g., Xcode, Swift, Objective-C) and filtered questions that resembled queries typically asked in customer support scenarios. We also excluded the questions that are too short (i.e., less than 100 characters) or too long (i.e., more than 800 characters) and contain screenshots. We matched the topic of all question-and-answer pairs to iPhone usage. We fixed basic grammatical errors or mistakes in all of the questions. Table~\ref{tab:questions} shows the seven questions we selected to use in our studies.

\subsection{Reference Documents for Accuracy Evaluation}
\label{sec:ref-doc}
For each question and answer pair, we created a reference document. We chose to provide the reference document for two reasons. First, if there is no reference document, participants would have to go through random materials on the internet --- some might look for official documentation, others might look for curated content like blog posts. Thus, adding the reference document helps control the performance difference caused by the search capability or preferences of the participants. This is useful because, in this paper, we are more interested in the performance difference caused by human-AI interactions rather than search capability or preference. Another reason is that the reference document is commonly used in a customer support environment, which is one of the representative question-answering applications in the real world. Someone answering customer questions about a company's product would likely have access to the documentation or manual for that product as well.

\begin{figure}[h]
    \caption{The detailed prompt provided to the AI agent for generating reference documents, including the title of the question and the body of the question, along with specific instructions for the document's format.}
    \label{fig:prompt-document}
    \centering\includegraphics[width=0.9\textwidth]{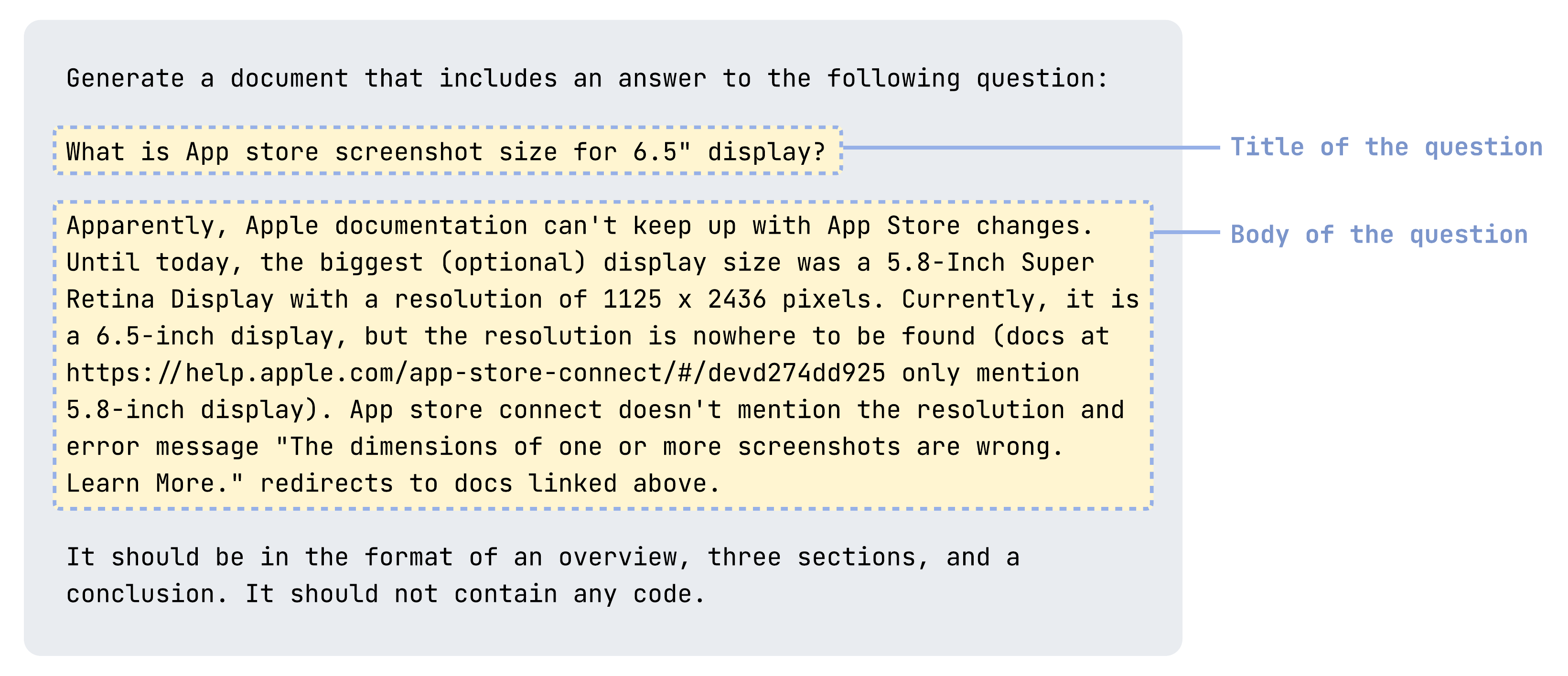}
\end{figure}

We used prompt engineering with \blue{GPT-4} to create high-quality reference documents  \cite{denny2023conversing, zamfirescu2023johnny}. Our initial prompt ``Generate a document that includes an answer to the following question'' led to two problems: the generated document was too short and it only contained the answer to the question and no other details which did not resemble typical specifications or reference documents \cite{reddy2022literature, roos2013customer}. Additionally, the generated document sometimes contained code. We assumed this is because the source of the questions is Stack Overflow, which is a platform heavily focused on programming. To include non-programmers in our study, we ensured the reference document did not generate code. We addressed the first issue of length by adding an instruction to our prompt: ``It should be in the format of an overview, three sections, and a conclusion''. We addressed the second issue by explicitly adding the instruction "It should not contain any code" at the end of the prompt. The resulting prompt we used for reference document generation is shown in Figure~\ref{fig:prompt-document}. The summary of the question used in the prompt was manually generated by the authors. 

The authors reviewed and corrected issues in the generated reference documents. For instance, since \blue{GPT-4} was refusing to give information about jailbreaking, the authors had to add more information about it in the generated document. Furthermore, we tried to minimize the overlaps between each reference document to avoid participants becoming familiar with the content. The authors independently reviewed each document for consistency and accuracy, ensuring that the correct answer to the question was included in the documents. The final documents we used were 3308.429 letters long on average ($SD = 608.020$).

\blue{The generated reference documents served two purposes in our experiments. First, they guided participants in constructing accurate responses during the formative study and Study 1. Study 1 instructions specified that correct answers could be found within the reference documents, ensuring participants consulted them rather than external sources. Second, the documents served as a standard for raters in Study 2 to evaluate response accuracy. While raters were given general instructions to "choose a better response," the reference documents provided an objective standard for accuracy, reflecting real-world problem-solving criteria. Findings from the formative study (\ref{sec:formative}) informed the design and methodology of Studies 1 and 2 (\ref{sec:design}, \ref{sec:methodology}).}

\section{Formative Study}
\label{sec:formative}
\blue{In the formative study, we explored various configurations to examine opportunities and challenges in human-AI collaboration for question answering. The results of this study informed the interaction designs in Study 1, enabling user-friendly configurations to address our research questions.}

\subsection{Participants}
We recruited 7 participants from Amazon Mechanical Turk and 5 participants from Prolific. All participants identified themselves as native or fluent in English. We screened out the participants who were only conversational or basic in English, considering that the task in our study requires at least a fluent level of English to be successfully completed. 6 participants graduated from college, and 6 participants completed graduate school.

\subsection{Study Conditions}
\blue{In the formative study, we tested four conditions (Conversation, AI Response, Highlight, and Similar Questions) designed to explore different levels of support in human-AI collaboration for question answering. Each condition builds upon the basic interaction of a human working with an AI agent (\blue{GPT-4}) and a reference document, adding layers of assistance to enhance the task. Below we list the conditions for the formative study in more detail.}

\begin{enumerate}
    \item \textbf{Conversation: Present the reference document \& Human interacts with the AI agent}: Participants interact with an AI agent and are provided with a reference document.
    \item \textbf{AI Response: Present the reference document \& Human interacts with the AI agent \& Provide AI-generated response as a starting point}: In addition to the AI agent and the reference document, participants are presented with an initial AI-generated response to the question to begin with.
    \item \textbf{Highlight: Present the reference document \& Human interacts with the AI agent \& Provide highlights of the reference document that are relevant to answer the question}: In addition to the AI agent and the reference document, participants are provided with the highlights of relevant information on the document.
    \item \textbf{Similar Questions: Present the reference document \& Human interacts with the AI agent \& Present nine similar questions with existing responses}: In addition to the AI agent and the reference document, participants can refer to nine similar questions and their existing responses, selected using Google Search results from Stack Overflow. 
\end{enumerate}

Each participant was exposed to all of the four conditions above, with the order randomized to control for order effects. Participants were given different questions for each condition to prevent memorization effects. Additionally, there was a short break between conditions to minimize fatigue.

\subsection{Apparatus and Procedure}
To reproduce the real-world situation, we utilized question-and-answer sets from Stack Overflow, one of the most popular question-answering platforms in the world. For each question and answer pair, we created reference documents. These reference documents served as the basis for participants to construct responses. Section \ref{sec:basic} elaborates on how these questions and documents were generated. 
Participants were provided a question and were asked to construct a response under their assigned conditions. In the post-survey, we asked participants questions about the task, such as "Which did you prefer the most among the four tasks?"

\subsection{Findings}

Participants reported that they found the condition that provides highlights of the reference document as the most helpful. Half of the participants preferred the \textit{Highlight} condition the most, 5 preferred the \textit{AI Response}, 1 participant preferred the \textit{Conversation}, and none preferred the \textit{Similar Questions} condition. The reasons included that the highlights were \textit{"helpful in identifying specific information needed to respond to the question"} and made it \textit{"easier to pinpoint the important information right away based on the question"}. No one reported the last configuration with 9 similar questions as preferable. Only one participant preferred the first condition with an AI agent without additional support. They mentioned, \textit{"the extras were a bit confusing."}



\blue{Many participants struggled to interact with the AI assistant, unsure of what to ask or what it could do. Notably, 8 out of 12 participants skipped using the AI in at least one condition, opting to write responses independently.}

\blue{Two participants interacted with the AI in the first condition but did not use it in subsequent conditions. The other participants used the AI in one or two conditions. Participants were informed upfront that correct answers could be found in the reference documents, meaning the task could be completed without AI interaction. However, the complete avoidance of AI by some participants—possibly to save time or reduce cognitive load—was unexpected.}

\blue{Participants resisted interacting with the AI agent and often struggled to use it effectively. Of the AI responses, 36.73\% were either unhelpful (e.g., \textit{"I'm sorry, but I need more information to provide an accurate answer."}, \textit{"Could you please provide more information about what kind of app you are looking for and what purpose it should serve?"}) or meaningless (e.g., \textit{"Yes, as an AI language model, I am capable of conversing with humans in a natural and human-like manner."}). This demonstrates the AI's difficulty in addressing questions requiring contextual knowledge or indirect references. These findings informed the design of configurations in the main study.}
\section{Design}
\label{sec:design}
Based on the formative study results, we refined the conversational AI and the study conditions. Here, we describe our process of modifying these components.

\subsection{Conversational AI for Question Answering}
\label{sec:ai}

\begin{figure}[h]
    \caption{The structured prompt used to provide the AI agent with a question and its corresponding reference documents, highlighting the title, body of the question, and relevant reference information.}
    \label{fig:prompt-feed}
    \centering\includegraphics[width=0.9\textwidth]{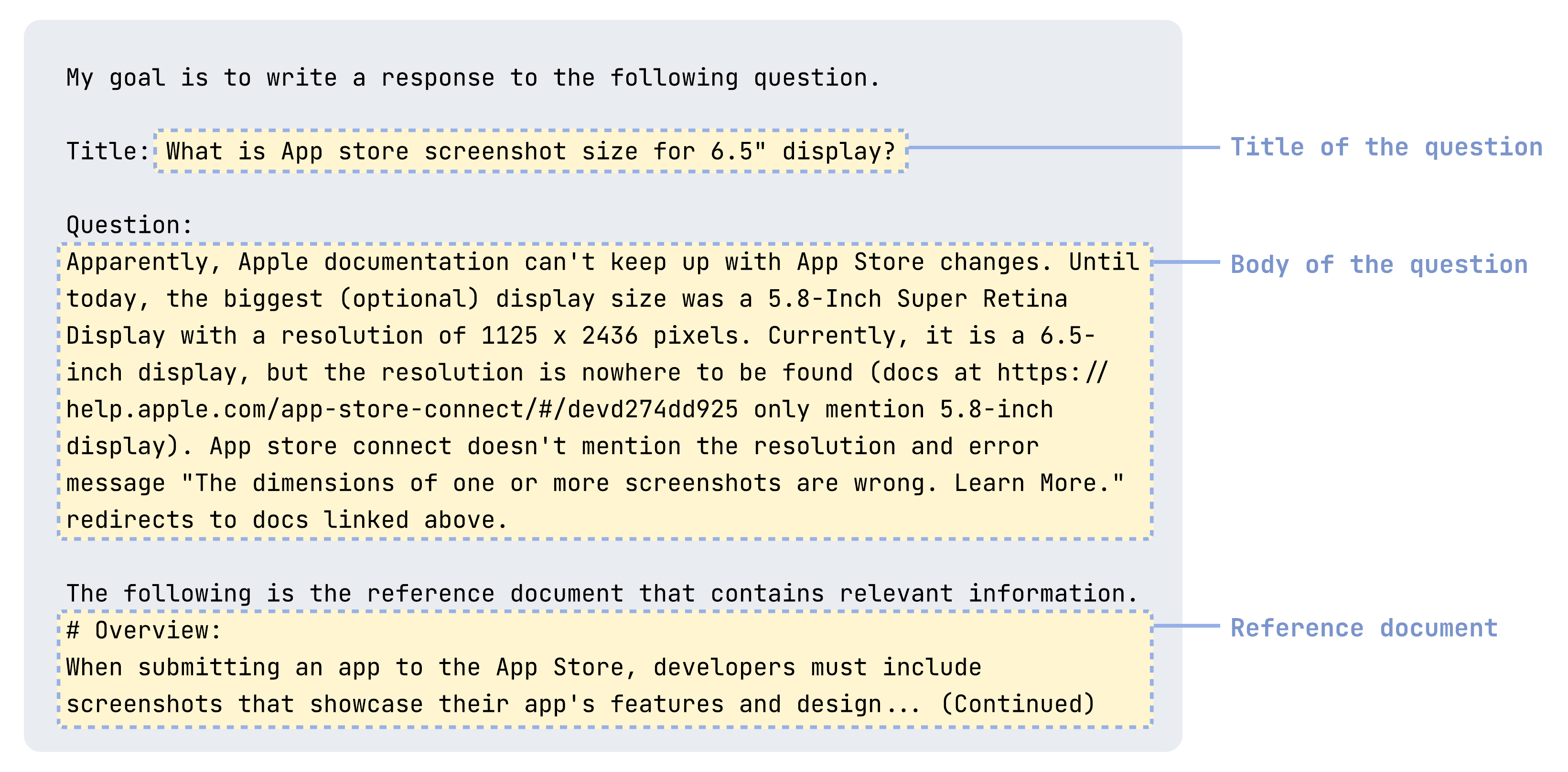}
\end{figure}

\blue{Conversational AI is widely used in question-answering across various industries. In our studies, we incorporated a conversational AI agent powered by \blue{GPT-4}. During the formative study, participants often assumed the AI already knew the reference document and question. To match these expectations, we provided the AI agent with the questions and documents before the experiments began. The prompt used for this setup is shown in Figure~\ref{fig:prompt-feed}. This preparation improved conversational continuity and created context to enhance human-AI interaction during the trials.}

\subsection{New and Refined Study Conditions}
Our formative study unveiled many shortcomings of our original conditions and how they had been designed. Our conditions for Study 1 were informed by the findings in the formative study. Below, we introduce a new condition, \textit{Nudging} and re-introduce the refined \textit{Highlight} condition. We provide our rationale, based on findings from the formative study, below. 

\subsubsection{The \textit{Nudging} Configuration: Query Shortcuts to AI Agent}

In the formative study, we found that a lot of participants had difficulties with how to best use the AI, which made them often not use it, or their interactions with the AI agent led to misunderstandings or breakdowns. Thus, we designed the \textit{Nudging} configuration that recommends three messages that can guide the participants to successfully get help from the AI assistant and initiate the conversation. For the \textit{Nudging} configuration, we added three buttons that a user could select from: ``Ask Agent what it can do to help'', ``Paraphrase the question'', and ``Summarize the reference document.'' \blue{We selected these three specific buttons based on two key criteria: their high frequency of use observed in our formative study and their alignment with tasks where AI demonstrates notable strengths, such as paraphrasing and summarization. These areas are well-suited to the capabilities of large language models.}

\begin{figure}[h]
    \caption{Comparison of the \textit{Nudging} and \textit{Highlight} configurations in AI Assistant design. \textit{Nudging} prompts users with additional cues to guide their interactions, while \textit{Highlight} emphasizes specific parts of the reference document to aid user understanding.}
    \label{fig:configuration_design}
    \centering
    \includegraphics[width=\textwidth]{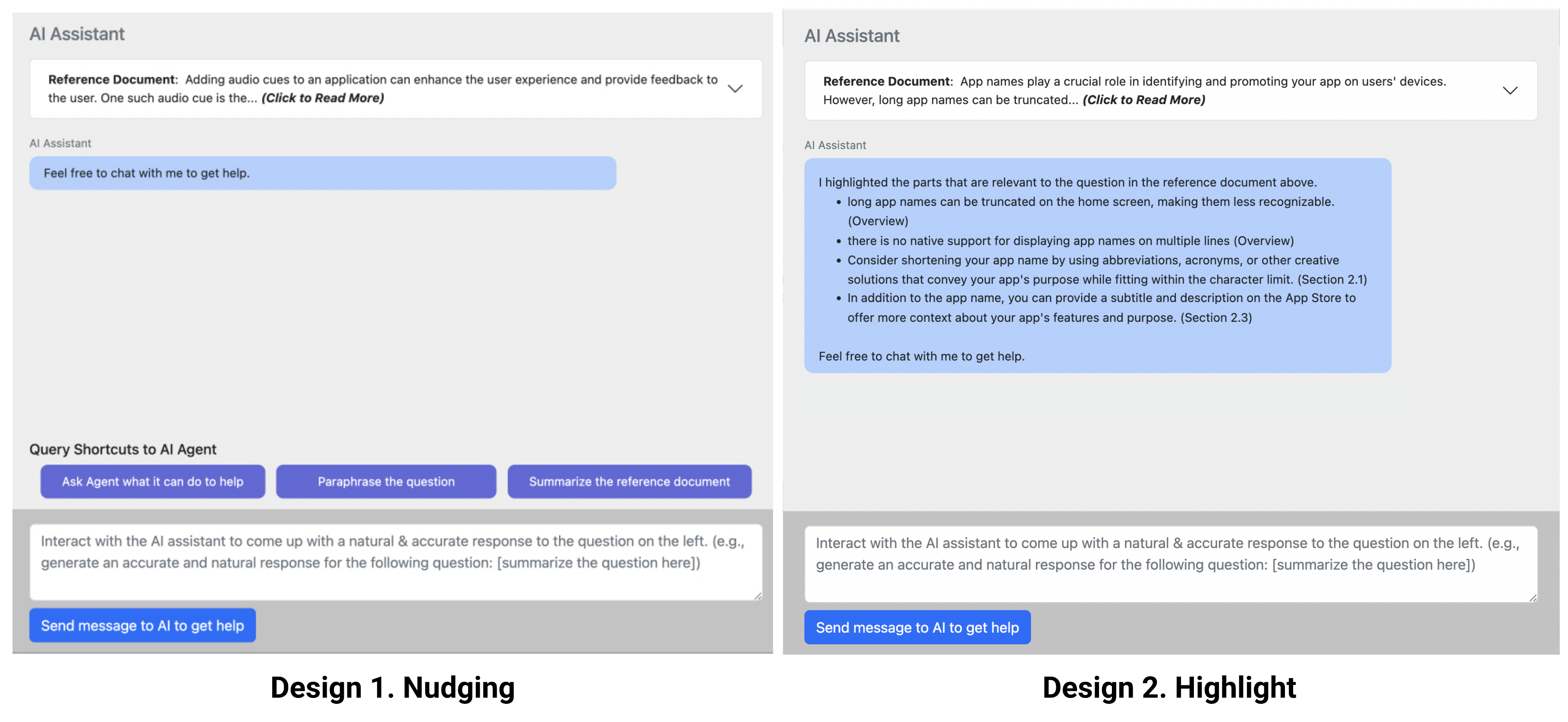}
\end{figure}

These suggested messages were presented as "query shortcuts to AI agent'' right on top of the input box where participants can send a message to the AI. The order of the recommendations was randomized to prevent order effects. Figure~\ref{fig:configuration_design} shows the screenshot of these buttons in the \textit{Nudging} configuration. When participants clicked on the button, the system inserted the corresponding full-length prompt in the input box. The full-length prompts were ``Could you please summarize the reference document?'', ``Can you please paraphrase this question for me?'', and ``What can I ask you to get help?'' respectively. These prompts in the input box were editable, so participants could additionally update them and then send them to the AI agent.

\subsubsection{The \textit{Highlight} Configuration: Presenting Key Sentences of the Reference Document}

The \textit{Highlight} configuration provides a list of key sentences highlighted from the reference document that can help answer the given question. This configuration was already tested in the formative study. We maintained this configuration in the main study because it was the most preferred condition in the formative study. The key sentences to highlight were extracted from the reference documents using \blue{GPT-4}. We made a few changes to the prompt design considering two issues identified from the formative study. In the formative study, participants gave feedback that few of the key sentences were irrelevant to answering the given question. For instance, the sentence ``Design an app icon that effectively communicates your app’s purpose and brand'' was highlighted, although it does not help answer the question ``Is there any way to display app names on multiple lines?'' Thus, we explicitly mentioned in the prompt that the key sentence should ``answer the question.'' Additionally, because the reference document was presented with highlights, some participants falsely believed that they could instruct the AI agent to highlight other parts of the documents, a feature that was not implemented, as it was a purely conversational AI. To prevent such confusion, we decided to present key sentences as a bullet-pointed list in the AI agent's initial message instead in lieu of highlighting them. The initial message from the AI agent listed the key sentences along with where they are located in the document.

\begin{figure}[h]
    \caption{The prompt used to generate highlights of the reference document}
    \label{fig:prompt-highlight}
    \centering\includegraphics[width=0.9\textwidth]{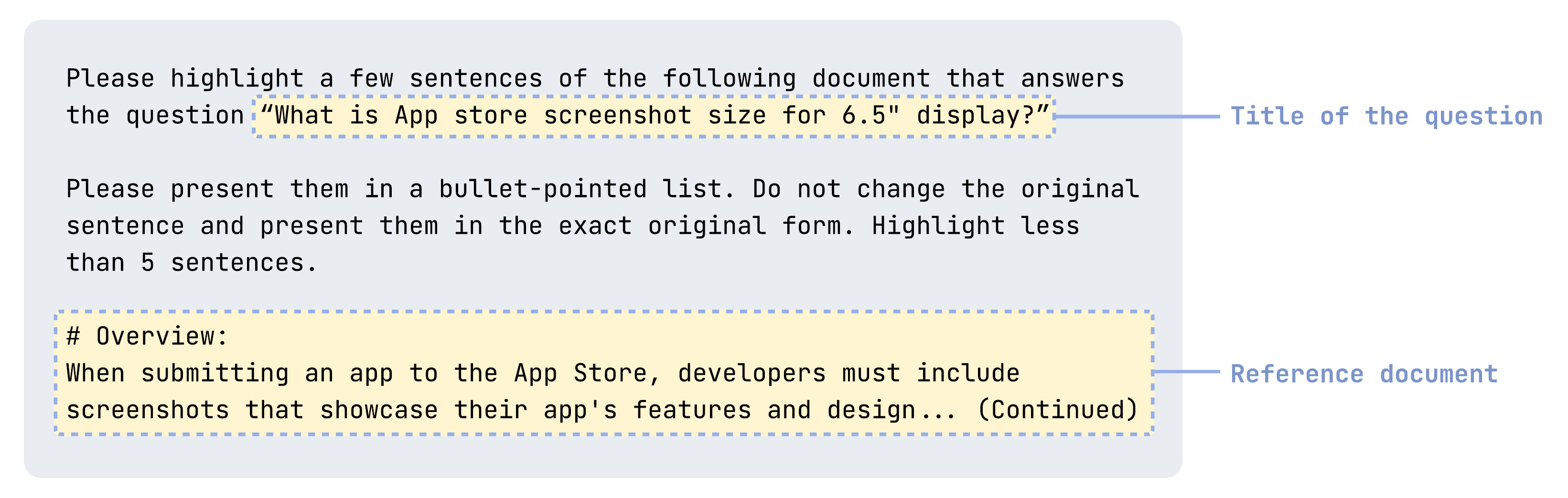}
\end{figure}

While we were further exploring the prompt engineering for generating highlights, we found that bullet-point lists suggested by \blue{GPT-4} were often too long. Accordingly, we added a sentence saying ``Highlight less than 5 sentences'' at the end of the prompt. We also found that \blue{GPT-4} often presents a paraphrased sentence that is different from the original sentence in the reference document. To prevent confusion, we added an explicit instruction for \blue{GPT-4} to keep the key sentences in the exact original form. The prompt can be seen in Figure~\ref{fig:prompt-highlight}. In the AI-generated results, the authors manually added the location of the highlighted sentences as shown in Figure~\ref{fig:configuration_design}.

\section{Methodology}
\label{sec:methodology}
Our main research consists of two studies (Study 1 and Study 2) aimed at evaluating and investigating the impact of different designs and interactive behaviors for question answering on the perceived quality of the responses generated. The studies are designed to address the following research questions:

\begin{itemize}
    \item RQ1. What kind of configurations overall lead to successful human-AI collaboration in question answering?
    \item RQ2. What interaction configurations help improve and generate high-quality responses in human-AI teams?
    \item RQ3. What are the characteristics of a good response in AI-assisted question answering?
\end{itemize}

The two studies are designed as follows:

\begin{enumerate}
    \item \textbf{Study 1}: Participants construct a response to a given question under varying conditions.
    \item \textbf{Study 2}: Raters evaluate the quality of the responses collected from Study 1 through pairwise comparisons.
\end{enumerate}

\subsection{Study 1: Response Construction}

\subsubsection{Participants}
 
Participants for Study 1 were recruited through Prolific, an online platform that specializes in participant recruitment for academic research \cite{palan2018prolific}. For this study, we aimed to recruit participants who are 18 years or older, speak English as their first language, have a minimum approval rate of 99, and have a minimum of 250 previous submissions. We specified screening criteria on Prolific to select participants who met these requirements. In addition, Prolific ensured that participants had a good track record of providing reliable responses in previous studies. This approach helped in recruiting a sample that is reliable for the research questions at hand. As a result, we hired 31 Prolific workers to participate in our study. Participants were compensated with a base payment of 4.50 USD for a 30-minute task. There was a bonus payment of up to 7.50 USD depending on the quality of their submitted responses, which was measured in Study 2. They were informed about the duration and nature of the study before enrolling, and were required to give informed consent.
\blue{We recruited 31 participants through Prolific, specifying criteria for eligibility: 18 years or older, native English speakers, a minimum 99\% approval rate, and at least 250 prior submissions. Prolific ensured participants had a strong track record of reliable responses. Participants were compensated with a base payment of 4.50 USD for a 30-minute task, with a potential bonus of up to 7.50 USD based on response quality measured in Study 2. All participants were informed about the study's duration and nature before enrolling and provided informed consent.}

\subsubsection{Study Conditions}

In this within-subjects design, each participant was exposed to all of the following conditions, with the order randomized to control for order effects:

\begin{enumerate}
    \item \textbf{Human: Present reference document}: Participants are provided with a reference document that includes an answer to the question.
    \item \textbf{Conversation: Present the reference document \& Human interacts with the AI agent}: Along with the reference document, participants can interact with an AI agent and get help from them in constructing the response.
    \item \textbf{Nudging: Present the reference document \& Human interacts with the AI agent \& Recommend several messages as guidance to prompt the AI agent}: Along with the reference document and AI agent, participants are provided with suggested messages to send to the AI.
    \item \textbf{Highlight: Present the reference document \& Human interacts with the AI agent \& Provide highlights of the reference document that are relevant to answer the question}: Along with the reference document and AI agent, participants are presented with a list of key sentences in the document.
\end{enumerate}

\begin{figure}[h]
    \caption{The prompt used to generate responses for the AI-only condition}
    \label{fig:prompt-ai}
    \centering\includegraphics[width=0.9\textwidth]{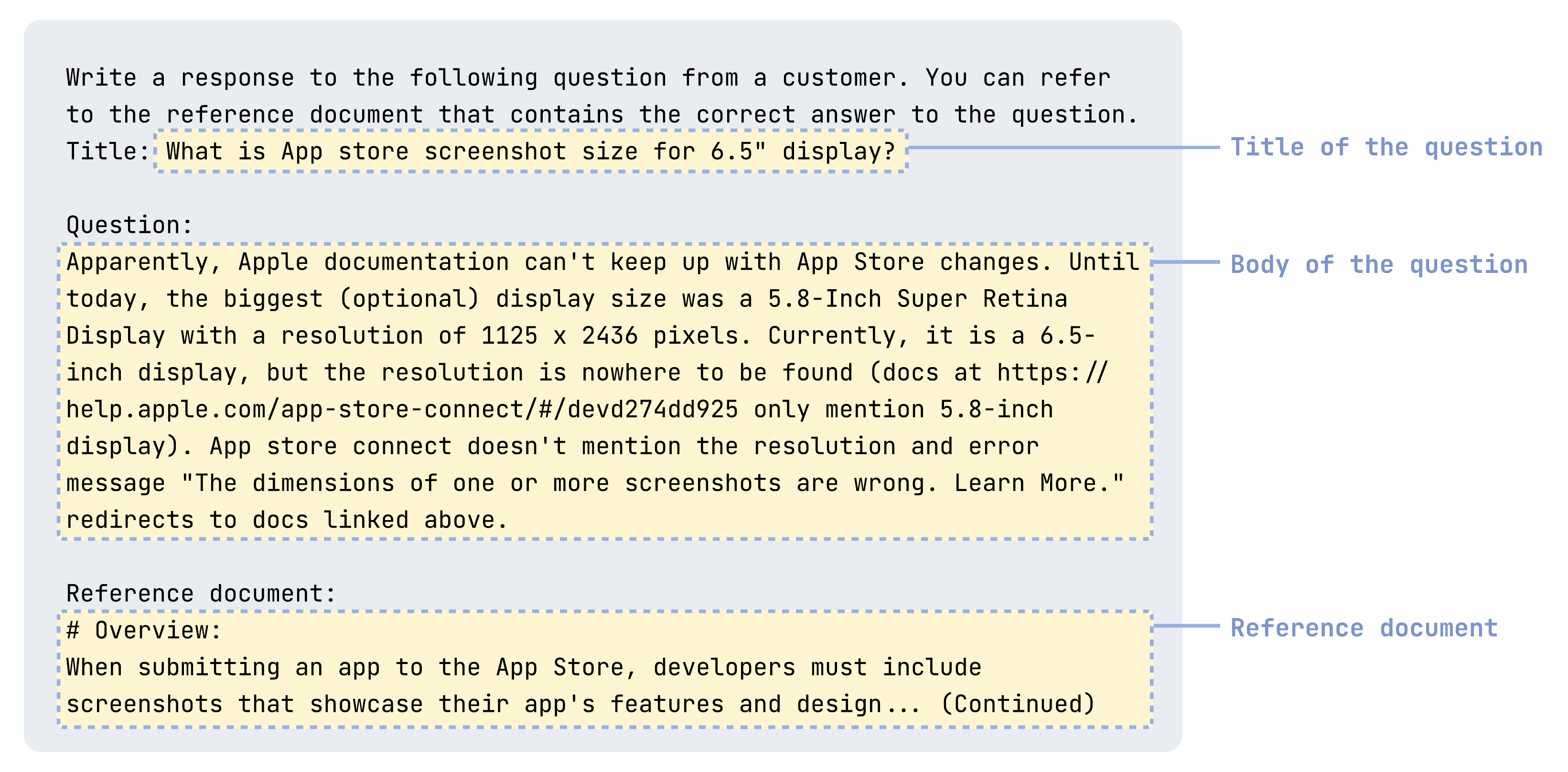}
\end{figure}

Participants were given a different question for each condition to prevent memorization effects. Additionally, there was a short break between conditions to minimize fatigue. We compared responses created from these four conditions to an AI-only response where \blue{GPT-4} generates the response. The AI-only condition did involve minimal human intervention by the authors. The responses for the AI condition were generated using \blue{GPT-4}, using the prompt in Figure~\ref{fig:prompt-ai}. However, we did not go through any iteration to improve the response and just used the initial response from the \blue{GPT-4} for the study.

\subsubsection{Apparatus and Procedure}

\begin{figure}[h]
    \caption{Experimental system interfaces used for each study condition}
    \label{fig:interface}
    \centering
    \includegraphics[width=\textwidth]{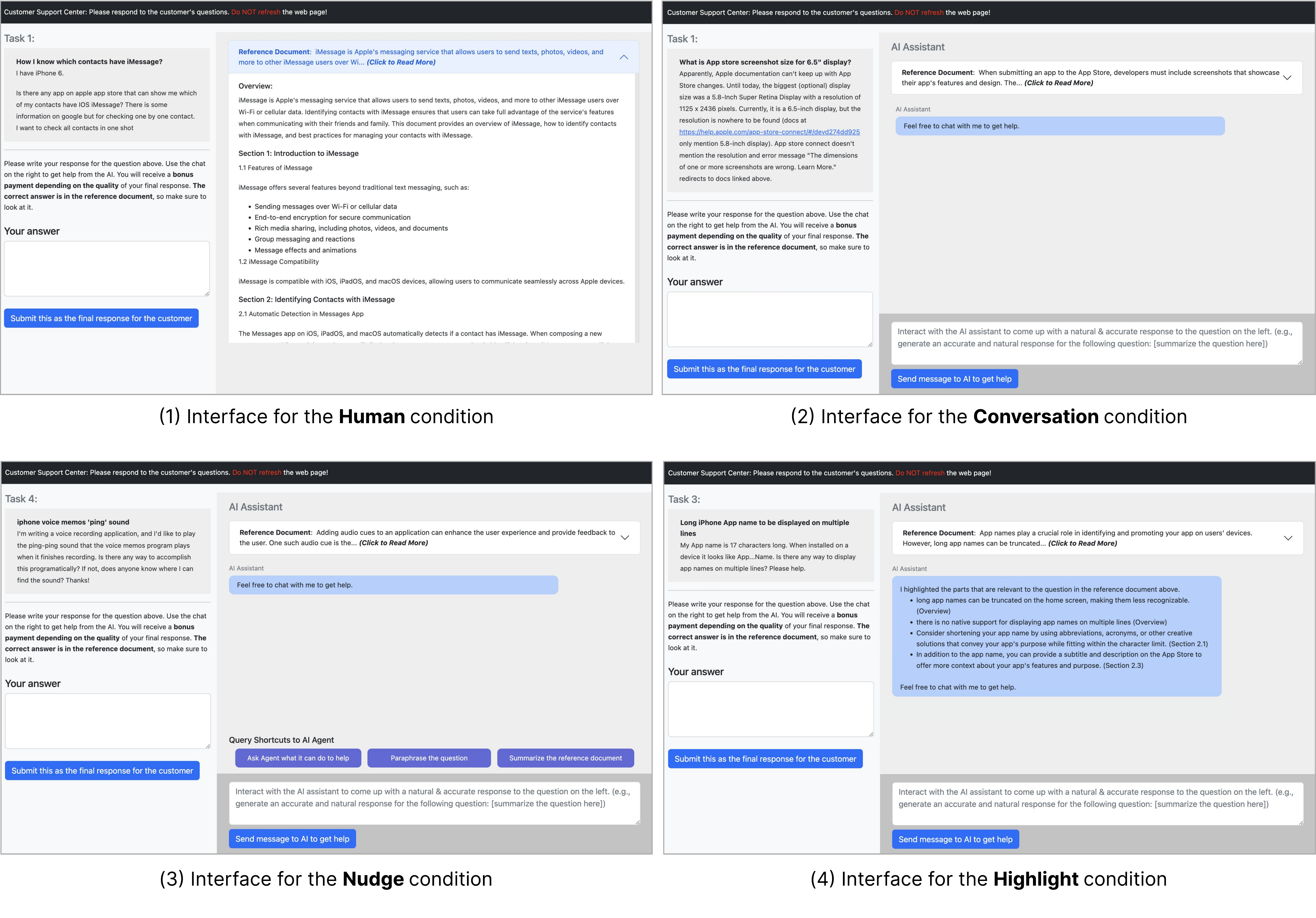}
\end{figure}

\blue{In the task, participants were presented with a randomly selected question from a set of seven Stack Overflow questions (listed in Table~\ref{tab:questions}). Each participant's question was randomly sampled, ensuring diverse question exposure across conditions. Participants then constructed their responses under their assigned condition, using a provided reference document as a criterion of accuracy.} They used the web-based system we provide as shown in Figure~\ref{fig:interface}. The responses generated by the participants, along with the time taken to construct the responses, and interaction logs with the AI agent (where applicable) were recorded. The experimental platform was implemented using React.js, Express.js, MongoDB, and OpenAI API for GPT-4. The system was loaded after the AI agent was internally prompted with the question and reference documents as described in Section~\ref{sec:ai}. The experiment took 73.277 minutes on average.

\subsubsection{Qualitative Data Analysis}


We collected task results such as the messages that participants exchanged with the AI assistant. These messages between humans and AI were subjected to content analysis. The goal was to understand the nature of the information provided by the AI, its relevance and contribution to the task, and the frequency and types of interventions or corrections made by human participants. Two authors separately open-coded the messages. The initial inter-rater reliability was assessed using Cohen’s Kappa ($\kappa$), yielding values of $\kappa = 0.766$ for AI messages and $\kappa = 0.638$ for human messages. Discrepancies in annotations between authors were discussed until a consensus was reached.


\subsubsection{Quantitative Data Analysis}

Quantitative data in Study 1 were collected across multiple categories including demographics, metadata, log-based usage of AI, and other log-based behaviors. Descriptive statistics were computed for each of the variables to provide an overview of the data. For continuous variables, measures such as mean, median, standard deviation, and range were calculated. For categorical variables, frequencies and percentages were computed. Inferential statistics were used to test hypotheses and to determine if differences between groups or conditions were statistically significant. The choice of inferential statistics was guided by the type of variable and the distribution of the data. We ran t-tests or ANOVA and Tukey's HSD tests to find out the difference in variables by study condition. In addition, we conducted correlation analysis to explore relationships between variables, such as the relationship between the quality of the response and the total number of messages asking for summarization.

\subsection{Study 2: Response Evaluation}

\subsubsection{Participants}

A separate group of 106 raters was recruited similarly to Study 1. For this study, we aimed to recruit raters who are 18 years or older, speak English as their first language, have a minimum approval rate of 99, and have a minimum of 350 previous submissions. We specified screening criteria on Prolific to select participants who met these requirements. Participants were compensated with 8.00 USD for a 50-minute task. Before starting, raters were instructed to use the reference documents for accuracy evaluation and shown examples of good and bad responses. An attention check during the task instructed raters to select a specific response. Those who failed were disqualified. Of the initial group, 39 failed or did not complete the task, leaving 106 raters who passed and finished all tasks.

Before starting the task, we directed raters to refer to the reference documents when evaluating the accuracy of the given responses. Then we showed three examples of responses that we regard as good or bad. Furthermore, there was an attention check in the middle of the task. In the attention check, the presented responses included an instruction saying "Make sure to choose Response 2 (Right). This is an attention check." If a rater failed to follow the instructions, they were no longer eligible to proceed to rate the responses. There were 39 people who failed the attention check or did not finish the task until the end. All 106 raters that we selected passed the attention check and finished all rating tasks.

\subsubsection{Apparatus and Procedure}


\red{There were 131 responses to be rated in total. Responses were sampled such that each question had an average of 19.714 responses (SD=3.239), achieved through a randomization process. Among these responses, 7 were from GPT-4, while the remaining 124 responses were generated by 31 participants in Study 1.} These responses resulted in 1,197 pairs in total, since only the responses to the same question can be paired together. Each rater was presented with 40 pairs of responses generated in Study 1, anonymized and without labels, indicating if the response was created by a human alone, AI alone, or human and AI together. The 40 pairs assigned to each rater were randomly sampled from pairs that had been rated fewer than three times to ensure a balanced review distribution. For each pair, raters evaluated which response was better. As a result, each pair was rated by about 3 raters. The collected ratings were used for our quantitative analysis to see correlations with other variables we collected in Study 1.

Prior work has shown that LLM may produce responses that users find appealing but that may ultimately lack factual accuracy~\cite{khurana2024and}. To address this, we took steps to ensure that raters considered factual accuracy alongside aspects like naturalness and user satisfaction in their evaluations. We explicitly highlighted these criteria in the rater instructions and provided a reference document as a benchmark for accuracy during the pairwise comparison task. The following is the full text of the criteria provided to the raters: ``Most importantly, you should evaluate each customer support response in terms of accuracy. The response should be \textit{accurate}. Fact-check the response using the reference document we provide you. The reference document contains the accurate information needed to answer the customer question provided. In addition to being accurate, the response should be \textit{natural}. Read the response carefully to see if the general tone is natural and appropriate. Overall, you should choose a response that is more likely to satisfy the customer who asked the question. We will bonus you if your ratings align well with the expert ratings.''

After completing the blind comparison of all pairs, raters were then presented with labels indicating the method through which each response was generated (i.e., human alone, AI alone, or human and AI together). They were asked to re-evaluate the responses, this time taking the labels into account. By checking the change in their evaluation after showing the source of the responses, we tried to examine the effect of biases toward AI or humans on the response evaluation. Raters were asked a series of survey questions to gain insights into their thought process and perceptions regarding the responses. The survey questions probed into the criteria used to determine which was the best response when presented with a pair of responses. The specific questions we asked are listed below. We performed open coding on the survey responses. 

\begin{enumerate}
   \item When you did not know the source of each response, what were the overall criteria you applied to choose one response over another?
   \item When you did know the source of each response, what were the overall criteria you applied to choose one response over another?
    \item What did you do when it was not clear which one was better?
\end{enumerate}



\section{Results}

\subsection{Overall Behaviors}

While participants were constructing the response to the provided question, there were 163 clicks on the reference document, 130 prompts to the AI assistant, and 123 responses from the AI assistant. Participants copied the reference document 76 times, copied messages from the AI assistant 66 times, and copied the question 17 times. They also pasted some text to the final response to submit 126 times and pasted some text to the prompt for the AI assistant 29 times. \blue{Overall, participants made a total of 308 edits, captured by a JavaScript blur event that triggers when an element loses focus. These interactions indicate frequent reference-checking and content adjustments throughout the task.} A total of 31 participants submitted one final response for each of the four study conditions, resulting in 124 responses. These 124 responses were rated by 106 raters via pairwise comparison. Using the ELO rating system, the average rating of the responses was set to 1500, consistent with prior work~\cite{elo1967proposed, pappalardo2019playerank}. Table~\ref{tab:responses} shows examples of highly-ranked responses and low-ranking responses.

We used content analysis to open-code the messages exchanged between participants and the AI assistant while performing the task. As a result, the AI messages were categorized into four groups: successful in responding to the participant's prompt, out of context, failed to respond, or refused to respond. The human messages were coded with the following categories: copying the question, copying the reference document, paraphrasing the question, asking the AI assistant to paraphrase or summarize something, asking a meta-level question (i.e., ``How can you help me answer this question?''), asking follow-up questions, asking AI to edit text, or others. By a follow-up question, we mean ``what is'' questions asked to the AI assistant (e.g. What is Cydia? What is an Android IDE?). These categories are specified in Table~\ref{tab:opencoding}. \blue{The presence of 9 human messages directly repeating the exact question to the AI agent, alongside 58 messages paraphrasing the question for the AI to generate an answer, implies that the human-AI collaborative conditions involved prompting the AI to generate responses. This approach mirrors how the AI-only condition works, where GPT-4 directly produces an answer to the question.}

\begin{center}
\begin{table}
\small
\caption{Examples of high-ranking responses and low-ranking responses.}
\label{tab:responses}
\begin{tabularx}{\textwidth}{|p{1.4cm}|p{1.7cm}|p{2.5cm}|p{6.85cm}|}
\hline
\textbf{Ranking} &\textbf{ Condition }& \textbf{Question Title} & \textbf{Answer} \\
\hline 

\multirow{2}{*}{High Rank}
 & Highlight&Long iPhone App name to be displayed on multiple lines&
\textit{Sadly longer app names are not supported on most platforms. App names are limited to 12-15 characters. depending on the device and the font size Longer app names like your present name will be truncated (as yours has been) My advice would be to consider a shorter app name that is both concise and descriptive but most importantly memorable and unique to you.} \newline\newline \textit{Please avoid using special characters and spaces in your app name as this can cause display issues on some devices and platforms. Please also comply with the guidelines of each platform. Otherwise the app may be rejected or removed from the store. }\newline\newline \textit{Consider using an app cons so it is easily recognized and then remembered by both users and potential new users. Finally in addition to app name you can provide a subtitle and description on the app store which will enable more context about your app's features and the purpose of the app. This can aid discovery and downloads.} \newline\newline \textit{Good luck } \\
\cline{2-4}
  & Conversation &In app purchase does not work when live&\textit{We are sorry to hear that your in-app purchase does not work when you are live. To start, please double-check that in-app purchases are correctly set up in App Store Connect and that their status shows as "approved" or "ready for sale." Once you have done that, please try restarting your app. If the issue persists, please move on to the next step.} \newline\newline \textit{I can see in your message that you said purchases were working in the sandbox environment, so please review the in-app purchase implementation in your app's code to ensure that purchases and transactions can be handled properly. Ensure that your app is fully up to date so that transactions can be handled smoothly. }\newline\newline \textit{Finally, please check your network connectivity to ensure there are no problems there, and also double-check that your app has not been flagged as non-compliance with the App Store guidelines.} \\
 \hline
 \multirow{2}{*}{Low Rank}
 & Highlight &Long iPhone App name to be displayed on multiple lines& \textit{You can abbreviate app names} \\
 \cline{2-4}
&Nudging&What is App store screenshot size for 6.5" display? &\textit{This is standard practice and the instructions are clear how to achieve it.} \\
 \hline
\end{tabularx}
\end{table}
\end{center}

\subsection{Quality of Responses by Study Condition}

\begin{figure}[h]
    \caption{Distribution of ratings before and after showing sources for various study conditions (AI, Human, Conversation, Nudging, Highlight). Overall ratings on average per condition (left) and top ratings on average for each condition (right)}
    \label{fig:ratings}
    \centering
    \includegraphics[width=\textwidth]{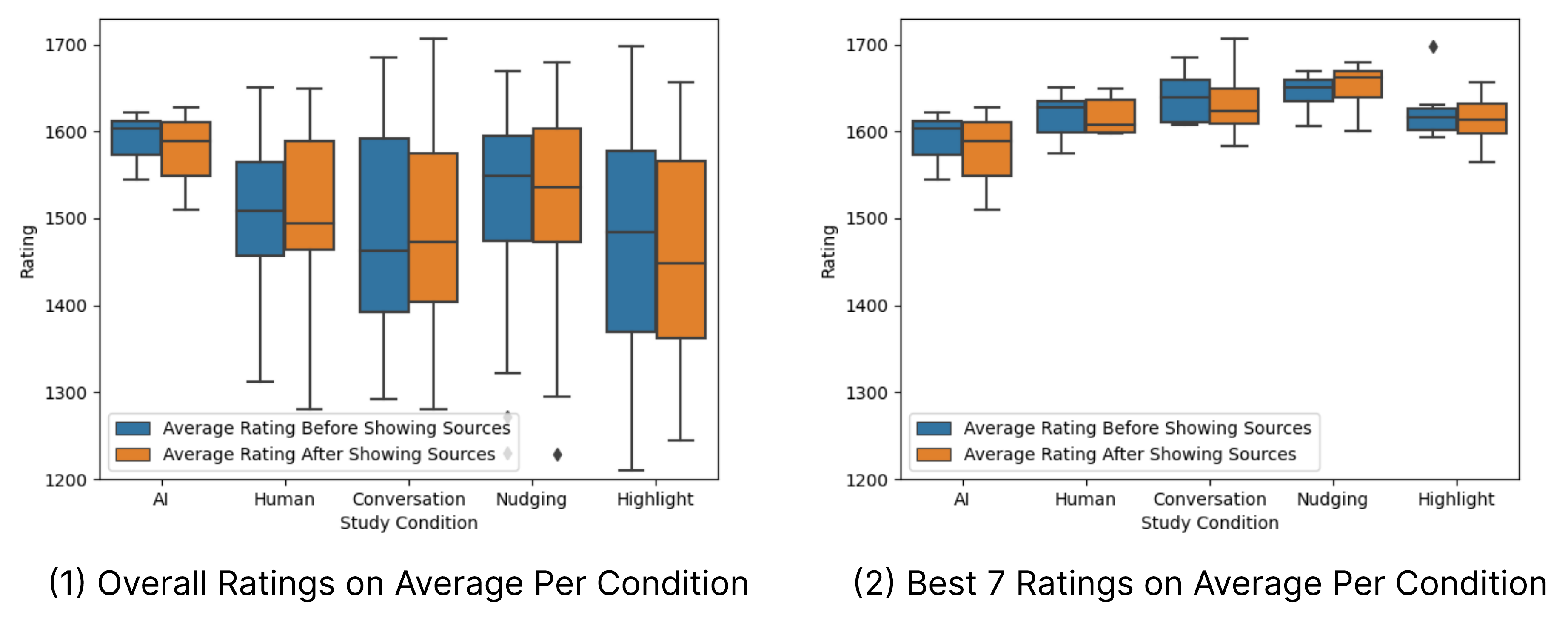}
\end{figure}


\blue{Figure~\ref{fig:ratings} shows a distribution of ratings per condition. While some variations were observed in average ratings before and after seeing the source, none of these differences were statistically significant for any of the conditions. For instance, the average rating for GPT-4 responses was slightly higher before seeing the source ($M=1591.730$) compared to after ($M=1577.818$), but this difference was not significant. The \textit{Nudging} condition was rated marginally lower before ($M=1524.825$) than after ($M=1525.888$), while the \textit{Human} condition showed a similar trend with lower ratings before ($M=1501.192$) than after ($M=1504.462$). The \textit{Conversation} condition had a slightly higher average rating before ($M=1486.312$) than after ($M=1485.683$), and the \textit{Highlight} condition was rated slightly higher before ($M=1466.958$) than after ($M=1466.395$). However, none of these differences were statistically significant.} 

\blue{To further investigate differences in average ratings across conditions, we conducted a one-way ANOVA. The analysis revealed significant differences in ratings before seeing the source ($p < 0.05$), but not after. Post-hoc comparisons using Tukey's HSD test indicated that GPT-4 responses were rated higher than those in the \textit{Highlight} condition before seeing the source ($p < 0.1$). However, no other pairwise comparisons between conditions showed meaningful differences either before or after the source was revealed.}


\blue{Meanwhile, the GPT-4 responses were produced through an optimized prompt engineering process designed to yield high-quality responses. To create a fair comparison, we aimed to optimize response quality in the other conditions as well, selecting the highest-rated interactions based on performance. Furthermore, there was a class imbalance across conditions: while the GPT-4 condition included only seven responses, one per question, the other conditions had over a hundred. By selecting the top seven responses in each condition based on their ELO ratings, we could address this imbalance, allowing for a balanced comparison of the highest-performing outputs across conditions. We used a method similar to how the class imbalance issue is usually addressed~\cite{hector2010analysis, kotsiantis2006handling}. This domain-specific approach to sampling the best responses enabled us to make meaningful comparisons between human, AI, and human-AI interactions on an equal footing.}


\blue{When considering only the best seven responses for each condition, small differences in average ratings were observed before and after seeing the source labels, but none of these differences were statistically significant. For instance, the \textit{Nudging} condition was rated slightly less favorably before ($M=1645.034$) than after ($M=1651.159$). Similarly, the \textit{Conversation} condition had higher ratings before ($M=1638.915$) than after ($M=1632.762$), and the \textit{Highlight} condition showed a higher average before ($M=1623.371$) than after ($M=1613.309$). The \textit{Human} condition was rated slightly less favorably before ($M=1616.975$) than after ($M=1618.171$). Finally, GPT-4 responses were rated more favorably before ($M=1591.730$) than after ($M=1577.818$).}

\blue{To examine differences in the average ratings of the best seven responses across conditions, we conducted a one-way ANOVA. The analysis revealed significant differences both before ($p < 0.05$) and after ($p < 0.01$) the source labels were shown. Post-hoc comparisons using Tukey's HSD test indicated that both the \textit{Nudging} and \textit{Conversation} conditions were rated significantly higher than \blue{GPT-4} responses, both before and after seeing the source labels ($p < 0.05$). No other pairwise comparisons between conditions showed statistically significant differences.}

Overall, the most successful human-AI collaborations are more useful than AI alone. These results suggest that human-AI collaboration is potentially useful, but only when their interaction is successful. The mere combination of humans and AI does not automatically guarantee performance improvement. The collaboration between them has to be successful in order to attain better performance.

\blue{We also conducted a two-way random-effects ANOVA to investigate the sources of the widespread distribution of ratings in human-involved conditions. Specifically, we examined two potential sources of variability: the individual participant who authored the response and the specific question that the response addressed. The analysis revealed that differences between questions did not significantly contribute to the variability in ratings, as indicated by a near-zero group variance estimate ($V = 0.026$). In contrast, the substantial residual variance ($R = 12490.9049$) indicates that most of the observed variation in ratings is likely attributable to differences between participants. This finding suggests that individual participants exhibit notable variability in the quality of their responses, which contributes to the wide spread in ratings.}

\subsection{What Leads to Successful Human-AI Collaboration? (RQ1)}


We found that certain factors were significantly correlated with the quality of the responses generated. Using the Pearson correlation coefficient, the ratings were positively correlated with the number of messages that asked the AI assistant to summarize text ($r = 0.310, p < 0.1$) and the total number of edits made ($r = 0.397, p < 0.05$). These correlations suggest that the resulting quality may be enhanced with more active human engagement in the collaboration. 


Two authors annotated the response messages from the AI assistant with four categories: successful in responding to the user's prompt, out of context, failed to respond, and refused to respond (See Table~\ref{tab:opencoding}). The latter three categories were grouped as unsuccessful responses. Multiple variables were significantly correlated with whether the AI response was successful or not. Using the Pearson correlation coefficient, results indicate that participants were more likely to get successful responses from AI when they copied ($r = 0.581, p < 0.01$) or paraphrased the question ($r = 0.522, p < 0.01$) to ask the AI. Further, the number of successful responses was positively correlated to the number of questions asking the AI what it can help with ($r = 0.693, p < 0.01$). However, AI often fails to understand overly paraphrased prompts. For instance, AI did not understand “What is the best workaround for this problem?” because it used the word “problem” instead of the word “question” that was initially used to feed the question to the AI model.

\vspace{1cm}
\small
\captionof{table}{Categorization of AI and Human Responses with Examples.}
\label{tab:opencoding}
\begin{tabularx}{\textwidth}{|l|p{2.5cm}|l|p{3.5cm}|X|}
\hline
Sender & Category & Count & Description & Example \\ 
\hline
\multirow{4}{*}{AI} & Successful & 115 & A message that has successfully responded to the request & Q. What is App store screenshot size for 6.5" display? \newline A. For a 6.5" display, the recommended App Store screenshot size is 1242 x 2688 pixels (portrait) or 2688 x 1242 pixels (landscape). \\ \cline{2-5} 
 & Out of context & 6 & A message that is out of the context of the request & Q. What is the best workaround for this problem? \newline A. I'm sorry, but you haven't provided any information about the problem you are referring to. Please provide more details so that I can help you find a suitable workaround. \\ \cline{2-5} 
 & Failed to respond & 4 & A message that failed to respond to the request & Q. What is the app Andrios? \newline A. As an AI language model, I don't have access to the entire internet and its constantly changing content. However, to my knowledge, there is no such app called "Andrios" available on official app stores for iOS or Android devices. It is possible that it may be a third-party app that can only be downloaded from unofficial sources, or it could be a misspelling of another app's name. \\ \cline{2-5} 
 & Refused to respond & 1 & A message that refused to respond to the request & Q. Can you give advice on the following question? \newline A. As an AI language model, I cannot recommend or encourage jailbreaking as it can void the warranty of your device and potentially introduce security risks. \\ \hline
\multirow{8}{*}{Human} & Copy question  & 9 & A message that exactly copies the question and asks it to the AI & Create a response for "My App name is 17 characters long. When installed on a device it looks like App...Name. Is there any way to display app names on multiple lines? Please help." \\ \cline{2-5} 
 & Copy reference & 1 & A message that copies a part of the reference document & Create an answer using the following document (reference document pasted here) \\ \cline{2-5} 
 & Paraphrased question & 58 & A message that paraphrases the question and asks it to the AI & Whats the most efficient way of exporting health data from my iphone automatically for python \\ \cline{2-5} 
 & Ask to paraphrase & 11 & A message that asks the AI to paraphrase a text & Can you please paraphrase this question for me? \\ \cline{2-5} 
 & Ask to summarize & 13 & A message that asks the AI to summarize a text & Can you summarize the reference document for me? \\ \cline{2-5} 
 & Ask meta & 8 & A message that asks the AI what it can do to help & What can I ask you to get help? \\ \cline{2-5} 
 & Ask follow-ups & 28 & A message that asks follow-up questions to the AI & What are the pros and cons of installing an android app on an iphone using Andrios? \\ \cline{2-5} 
 & Ask to edit text & 4 & A message that asks the AI to edit a text & Can you write it in a more pleasant way \\ \hline
\end{tabularx}




\subsection{The Query Shortcuts Helped Improve the Interaction (RQ2)}


The query shortcuts in the \textit{Nudging} study condition were shown to improve the overall interaction. Results indicate that the message suggestions actually influenced and changed the participants' behaviors. For instance, more participants asked the AI to paraphrase the reference document in the third condition than in other conditions. Using Tukey's HSD test, the number of messages asking the AI to summarize a text was significantly greater in the \textit{Nudging} condition where there were suggestions than in the \textit{Highlight} condition ($p < 0.01$). The number of messages asking the AI assistant to paraphrase a text was also significantly greater in the \textit{Nudging} condition than in the \textit{Highlight} condition ($p < 0.01$) or \textit{Conversation} condition ($p < 0.01$). Furthermore, the number of messages asking what the AI can do to help them was significantly more frequent in the \textit{Nudging} condition than in the \textit{Highlight} condition ($p < 0.01$) or the \textit{Conversation} condition ($p < 0.01$). The total number of messages sent to the AI assistant was significantly greater in the \textit{Nudging} condition than in the \textit{Highlight} condition ($p < 0.01$).

Results show that these changes were positive in terms of human-AI collaboration and led to successful interactions. There were more successful responses from the AI assistant in the \textit{Nudging} condition compared to the \textit{Highlight} condition ($p < 0.01$) or the \textit{Conversation} condition ($p < 0.1$). Also, the total number of clicks on the message suggestions was positively correlated with the number of successful responses from the AI assistant ($r = 0.626, p < 0.01$), using the Pearson correlation coefficient.

\blue{In the survey, one participant commented on the \textit{Nudging} condition, saying, ``This helped me understand which features the AI had and which may be the best way the AI could help me.'' Another participant noted, ``It gave me a good starting point.'' These quotes suggest that the \textit{Nudging} condition effectively enhanced participants' awareness of the AI's capabilities and offered guidance on how best to leverage its assistance.}

\subsection{What Responses Are Perceived as Successful? (RQ3)}



We conducted open coding on the raters' responses to the survey question asking what their overall criteria were for choosing one response over another. As a result, their responses were annotated with seven different labels. The frequency of each label is shown in Table~\ref{tab:freq-rater}. The most frequently mentioned aspect is accuracy, which involves fact-checking the responses, whether they concur with the reference documents and are actually helpful in resolving the problem. The next frequent criterion is the tone of the response, checking if it is friendly, natural, and human-like. The third most frequent criterion is the length of the response. There was a disagreement about whether shorter or longer responses are desirable, which is elaborated on in Section~\ref{sec:length}. Participants also mentioned clarity, relevance, and grammatical correctness. Lastly, a few raters mentioned the ethical soundness of the response, due to the last question in Table~\ref{tab:questions} that mentions jailbreaking a phone.


\begin{table}[h]
\small
\caption{Frequency and Examples of Criteria Used by Raters to Evaluate Responses}
\label{tab:freq-rater}
\begin{tabularx}{\textwidth}{|l|l|X|}
\hline
Criteria & Frequency & Survey Response  \\ 
\hline
Accuracy & 58 & \textit{Firstly, accuracy of the information supplied according to the reference material; then the level of detail/inclusion of information such that it was an appropriate and useful response to the question; then the degree to which the response had a natural/friendly tone; finally minor issues such as formatting/typos/etc.} \\
\hline
Tone & 42 & \textit{Accuracy of response, natural sounding but not too patronising in tone and answering the question concisely without unnecessary waffle} \\
\hline
Length & 41 & \textit{If it sounded more humane than the other response and the response was lengthy enough to provide a suitable answer, and successfully did that} \\
\hline
Clarity & 24 & \textit{Clarity and accuracy of information provided. If both provided correct information, I tended to choose the response which used shorter, clearer sentences, and correct grammatical constructions.} \\
\hline
Relevance & 20 & \textit{I evaluated how extensive each response was, as well as how relevant it was. Some answers,although very long and seemingly detailed, went off topic and gave irrelevant information. }\\
\hline
Grammar & 17 & \textit{If it was coherent. Some of the text had poor grammar so I would use that information to make my decision.} \\
\hline
Ethics & 5 & \textit{That the response was not encouraging illegal jail breaking and that it had no grammatical errors/typos }\\
\hline

\end{tabularx}
\end{table}
\section{Discussion}

\subsection{What Do People Prefer: Human+AI-Generated Text vs. Human-Generated Text vs. AI-Generated text}

In total, considering all responses in each condition, raters preferred AI-generated text with an average rating of 1591.730. However, when we considered the best 7 ratings, they preferred human+AI-generated responses. Similarly to these pairwise comparison results, raters \textit{perceived} that the human+AI-generated responses were the most preferable responses, when asked for their overall thoughts about each type of response. They said the responses by human-AI combination are ``balanced in terms of factualness and naturalness''


\subsubsection{Effects of Showing the Author of the Response}

Our findings revealed that none of the conditions showed a significant difference in ratings given to the responses between those who showed the source of the response and those who did not. However, the post-survey results show that there exists a bias regarding who wrote the response. When asked how they rated different responses (with/without source), participants said ``I was still comparing the responses to see if they were accurate, well-written, and comprehensive.'' Several raters answered they prioritized human-generated responses. They mentioned, ``I accepted that the human working alone might have better information than the AI" or "If it was from a human and if it came across correct.'' This suggests that some raters tend to think that human-generated responses are superior to AI-generated responses. This supports prior work \cite{ashktorab2020human} that humans tend to be biased towards humans.

\subsection{Design Recommendations for Conversational AI}
Based on our findings, we are able to make recommendations on how to design for successful human-AI collaboration. In this section we make three recommendations grounded in our findings: (1) Guide the user on how to interact with the AI Agent, (2) Meta Prompt and (3) Ensure that the user and AI utilize shared vocabulary.

\subsubsection{Guided Message Suggestions Can Enhance Human-AI Collaboration}

Recommending messages that guide the conversation with AI and providing shortcuts to using these messages can improve human-AI interaction. For instance, the \textit{Nudging} condition had more successful responses from the AI than the \textit{Highlight} or \textit{Conversation} conditions ($p < 0.01$). Furthermore, the total number of clicks on the message suggestions was positively correlated with the number of successful responses from the AI assistant ($r = 0.626, p < 0.01$). These findings prove that the guidance on what kind of messages to send to AI helps achieve successful human-AI collaboration. 

To effectively support users in human-AI collaboration, query shortcuts should offer actionable prompts tailored to tasks where AI excels. This approach is particularly beneficial for users unfamiliar with AI tools, as it helps them initiate conversations more naturally and constructively. Future research could investigate adaptive and dynamic message suggestions that respond to timing and contextual cues, offering customized guidance that evolves throughout the interaction.

\red{A key caveat is that overly structured guidance may reduce user autonomy and creativity in forming queries, potentially leading to over-reliance on system-suggested prompts rather than organically exploring AI’s capabilities. Striking a balance between structured guidance and open-ended flexibility remains a challenge in designing effective AI-assisted interactions.}


\subsubsection{Meta Prompting in Human-AI Interaction: Ask AI What to Ask AI}

One of the available nudges in our \textit{Nudge} condition asked the AI agent how it should interact with the AI. In our study, this kind of meta-prompting was found to be positively correlated with the number of successful AI responses ($r = 0.693, p < 0.01$). Our formative study demonstrated that participants do not know how to interact with the AI agent or what to ask. Providing an option that allows them to ask the AI agent how it can help them is a strategy that our participants used to yield successful interactions.  Thus, the idea of asking the AI what it can help with can help enhance the human-AI interaction.

\subsubsection{Establishing Common Vocabulary for User-AI Interaction}

Although paraphrasing a question was positively correlated with the number of successful AI responses ($r = 0.522, p < 0.01$), paraphrasing sometimes caused failures of AI. For example, AI did not understand ``What is the best workaround for this problem?” because it used the word “problem'' instead of the word ``question'' that was initially used to feed the question to the AI. Thus, the AI agent should be designed in a way that makes sure it can understand synonyms and paraphrased versions of important keywords. To address shared vocabulary, we prompted our AI with components of the task (reference document, question) so it had an understanding of what these components were if the user referenced them. What we failed to do however was identify synonyms for ``question'' (problem, inquiry, challenge) not anticipating that users may refer to the question in different ways. One way to account for this behavior is to prompt the model with alternative ways of referring to the question or other components of the task. There were 6 messages marked as such a failure caused by a lack of paraphrasing capability among 126 messages in total (4.762\% failure rate). 

\subsection{Gaps between Subjective Preferences and Rankings of Responses}
\label{sec:length}
 In our study, we found a misalignment between the participants' reported preference and actual preference.  In the survey, raters expressed their preferences about the length of a response. 19 raters explicitly mentioned that they preferred short and concise responses, whereas 18 raters mentioned that they preferred long and detailed responses, an almost even split. 2 raters mentioned that they considered length, but there was a tradeoff between long and short responses. One of these two raters said ``Sometimes I went with answers that provided more info, other times with ones that were more concise: that varied depending on the content.'' However, we found that response length was positively correlated to the quality ratings ($r = 0.534, p < 0.01$). In other words, while the subjective preferences of response length were varied, in practice, the longer the response was, the higher participants rated it. 


We also found a discrepancy between participants' preferences for the source of the responses (human, AI, human+AI) and their actual ratings of those responses. Our survey findings that probed in user perceptions, show that the majority of raters perceived responses by human-AI combination as most preferable. A majority of the raters (71 out of 106) mentioned they thought the human-AI combination was the best, with only 8 raters selecting AI alone. The actual ratings participants gave when rating the responses, demonstrate a gap from these perceptions. Although many raters pointed out how AI responses were problematic, the average rating of the AI responses was higher than the average for all conditions. This indicates a potential bias that the raters tend to have on the AI-generated content. Despite being biased against the AI-only generated content when told how it was generated, they rated it most highly.

\subsection{Limitations and Future Work}


Our work has certain limitations which we acknowledge. 

One limitation is that the participants were not experts on iPhone usage, so the quality of the responses might have been worse than those written by experts. Furthermore, our study is limited in that we examine only certain kinds of context of question answering. For example, there could be question-answering scenarios where there is no answer in existing documents or materials. However, in our study, we mainly focused on situations where an answer exists somewhere in the reference document. Although this limits our study from being extended to questions without existing answers, this assumption is still useful for the purpose of experimentation because it provides a way to objectively evaluate the accuracy of the collected responses. 

Another notable limitation is that we synthetically generated the reference documents using \blue{GPT-4}. This approach was chosen for several reasons. Firstly, it allowed us to maintain strict control over the quality and formatting of the reference documents, ensuring consistency across all tasks. Secondly, in some cases, reference documents for the specific questions we utilized did not exist beforehand, necessitating their creation. It is important to note that the generation of these reference documents was a collaborative effort involving both humans and AI. Authors performed multiple iterations of prompt engineering and addressed issues in the generated documents which are documented in Section \ref{sec:ref-doc}. This approach using machine-generated synthetic data aligns with the broader research goal of improving privacy and enhancing the representativeness of study results, as discussed in \cite{syntehtic_2023}.

Our study is also limited in terms of scope and generalizability, as it evaluates only two configurations with a single LLM (GPT-4) and concentrates on three specific message suggestions. While we recognize this limitation, we believe our choices have value as an initial step toward a focused understanding of human-AI interaction support within the selected context. There is an inherent tradeoff between broadening and narrowing the study's scope. Expanding to include additional models, configurations, and message suggestions could enrich the generalizability of our findings. However, a narrower focus enables a more in-depth examination of specific configurations and their impact on performance. GPT-4, being a widely adopted model comparable to other consumer-accessible LLMs, provides a robust foundation for such focused exploration. \red{Additionally, we acknowledge that the rapid advancement of generative AI models may render certain findings less relevant over time, as newer iterations with improved capabilities and prompt configurations could address some of the challenges we explore.} Furthermore, the design of our two configurations and three message suggestions is grounded in findings from our formative study, lending relevance and specificity to our approach.

Lastly, our top-7 response analysis highlights a limitation: in practice, identifying high-quality responses before a full evaluation can be challenging. Filtering for high-skilled participants could be an alternative way to yield top-tier responses, as response variability reflects individual human capabilities. The top-k analysis also serves to showcase each condition’s potential to produce high-quality outputs, which is particularly valuable in contexts where a high-performing subset is critical. Still, we acknowledge that the top-k results can be limited in terms of practical utility.
\section{Conclusion}

In this paper, we discuss the challenges and complexities of human-AI collaboration in question answering. The study explores various interaction configurations and assesses their impact on response quality. Two configurations, \textit{Nudging} and \textit{Highlight}, are designed based on formative study findings. The results of controlled experiments with 137 crowd workers show that employing human-AI collaboration teams to answer questions can be beneficial only when the interaction between humans and AI is successful. The query shortcuts in the \textit{Nudging} configuration were shown to help enhance the interactions and lead to higher-rated responses. This paper further contributes to the human-AI interaction literature by providing insights into what factors can make AI-assisted question-answering effective.

\nocite{*}
\bibliographystyle{ACM-Reference-Format}
\bibliography{refs}


\end{document}